\documentclass[fonts]{icst}

\usepackage{moreverb}

\usepackage[breaklinks,colorlinks,bookmarksopen,bookmarksnumbered,linkcolor=ICSTblue,citecolor=blue,urlcolor=ICSTblue]{hyperref}
\usepackage{breakurl}
\usepackage{doi}
\usepackage{amsmath}
\usepackage[linesnumbered,ruled,vlined]{algorithm2e}
\usepackage{algpseudocode}
\usepackage{multirow}
\usepackage{multicol}
\usepackage{float}
\usepackage{graphicx}
\usepackage{tabularx}
\usepackage{lipsum}

\newcommand\BibTeX{{\rmfamily B\kern-.05em \textsc{i\kern-.025em b}\kern-.08em
T\kern-.1667em\lower.7ex\hbox{E}\kern-.125emX}}

\def\volumeyear{YYYY}

\journalname{XXXXXX}
\articletype{Research Article/Editorial}
  \def\copyrightholder{Author Name\textit{ et al.}, licensed to EAI}
  \copyrightnote{This is an open access article distributed under the terms of the \href{https://creativecommons.org/licenses/by-nc-sa/4.0/}{CC BY-NC-SA 4.0}, which permits copying, redistributing, remixing, transformation, and building upon the material in any medium so long as the original work is properly cited.}
  \def\articledoi{10.4108/XX.X.X.XX}
\received{XXXX}
  \accepted{XXXX}
  \published{XXXX}

\begin{document}

\runningheads{T. C. Minh et al.}{A demonstration of the \LaTeXe \ class file for \journalabb\ }

\title{UGGNet: Bridging U-Net and VGG for Advanced Breast Cancer Diagnosis}

\author{Tran Cao Minh\affil{1}, Nguyen Kim Quoc\affil{1}, Phan Cong Vinh\affil{1}, Dang Nhu Phu\affil{1}, Vuong Xuan Chi\affil{1}, \\ Ha Minh Tan\affil{1}\fnoteref{1}}
\address{\affilnum{1}Faculty of Information Technology Nguyen Tat Thanh University, Ho Chi Minh City, Vietnam}

\abstract{In the field of medical imaging, breast ultrasound has emerged as a crucial diagnostic tool for early detection of breast cancer. However, the accuracy of diagnosing the location of the affected area and the extent of the disease depends on the experience of the physician. In this paper, we propose a novel model called UGGNet, combining the power of the U-Net and VGG architectures to enhance the performance of breast ultrasound image analysis. The U-Net component of the model helps accurately segment the lesions, while the VGG component utilizes deep convolutional layers to extract features. The fusion of these two architectures in UGGNet aims to optimize both segmentation and feature representation, providing a comprehensive solution for accurate diagnosis in breast ultrasound images. Experimental results have demonstrated that the UGGNet model achieves  a notable accuracy of 78.2\% on the "Breast Ultrasound Images Dataset."}



\fnotetext[1]{Corresponding author.  Email: \email{hmtan@ntt.edu.vn}}

\maketitle

\section{Introduction}
    Breast cancer is a type of cancer originating from the cells of the breast, often arising from the milk duct cells or surrounding cells. This is one of the most common types of cancer in women worldwide \cite{jabeen2022breast, giaquinto2022breast}. To examine and detect the disease early, ultrasound imaging can be employed for diagnosis and monitoring of breast cancer. Ultrasound procedures can help determine the size of the tumor, and its characteristics, and identify suitable treatment options \cite{man2022role}. Based on the ultrasound images, the doctor remains the direct diagnostic authority regarding the patient's illness severity. There are two potential scenarios that a patient may face upon diagnosis: the ailment is either benign or malignant\cite{faust2018comparative}. In recent years, deep learning models have demonstrated outstanding capabilities in predicting and classifying diseases, particularly when employing convolutional techniques \cite{kumar2017convolutional, chattopadhyay2022mri}. This is considered a crucial method in image information processing. The application of convolutions has expanded into various medical fields, with the advantage of efficiently processing large image and data sets \cite{ghrabat2022fully, zahoor2022breast}. In the domain of disease diagnosis, particularly in breast cancer, convolutional techniques have proven to be effective.
    In particular, research on cancer diagnosis utilizing medical imaging is a promising area, as it has been deployed to predict diseases based on the progression of patients over time, employing a self-attention-based model proposed by Aishik Konwer et al., incorporating a Temporal Convolutional Network (TCN) \cite{konwer2022temporal}. Manu Subramoniam et al. have suggested the application of the Resnet model to predict Alzheimer's disease from MRI images \cite{subramoniam2022deep}. Additionally, Ahmet Solak et al. have proposed the use of a U-Net model for segmenting tumor masses within the adrenal gland \cite{solak2023adrenal}.
    
    In this paper, we propose the architecture of UGGNet and present experimental results on the BUSI dataset. Concurrently, this research discusses prospects and challenges in the future development direction of the topic, raising crucial issues that need to be addressed to enhance the efficiency and practical application of UGGNet in the field of medical diagnosis, specifically within the realms of computer vision and deep learning.
\section{Literature Review}
    Since the advent of the Backpropagation algorithm by Professor Geoffrey Hinton in 1986 \cite{rumelhart1986learning}, research applying deep learning (DL) to various aspects of life and medicine has been burgeoning almost daily. Breast cancer, considered a typical disease affecting women, has become a crucial and promising research area. With the advancement of DL, machine learning and deep learning models have been developed to handle complex medical data and provide accurate predictions of patients' health conditions. Mesut TOĞAÇAR et al. \cite{tougaccar2018deep} applied the Support Vector Machine (SVM) model to train on a dataset of 700 images, including benign and malignant variants of breast cancer images. The images were analyzed using a convolutional neural network (CNN) approach, and the input images were passed through the AlexNet model \cite{krizhevsky2012imagenet} for feature extraction. The extracted features were then combined with the SVM model for classification, achieving an impressive accuracy of 0.934. Ashutosh Kumar Dubey et al. \cite{dubey2016analysis} utilized the Breast Cancer Wisconsin (BCW) dataset, a structured dataset derived from digitized images of tumors. They described the characteristics of cell nuclei within the tumors and applied the K-means clustering algorithm, involving cluster initialization, distance measurement to the nearest clusters, and cluster optimization. The results yielded an average accuracy of 0.92 in experimental trials. Using the same BCW dataset, Omar Ibrahim Obaid et al. \cite{obaid2018evaluating} experimented with SVM, K-nearest neighbors (KNN), and Decision tree models, achieving an accuracy of 0.981. Notably, the Quadratic SVM kernel \{$K\left(x,y\right)=\left(x\cdot y+c\right)^d$\} demonstrated superior performance when dealing with complex relationships that cannot be well-classified by a simple linear boundary. Ensemble models have also emerged as a new approach. Mohamed Hosni et al. \cite{hosni2019reviewing} combined three individual models, Decision tree + SVM + Artificial Neural Network (ANN), to create a comprehensive strength. Abien Fred M. Agarap \cite{agarap2018breast} contributed a Multi-layer Perceptron (MLP) model, optimized the hyperparameters, and achieved an outstanding accuracy of 0.99, demonstrating the effectiveness of deep learning models.

    In recent years, the "Dataset of breast ultrasound images - BUSI" \cite{al2020dataset} has emerged as a benchmark for researchers to experiment with their models and compare results. Michal Byra et al. \cite{byra2020breast} proposed a segmentation model based on U-Net \cite{ronneberger2015u} combined with Selective Kernel (SK), achieving a Dice score of 0.778. Jorge F. Lazo et al. \cite{lazo2020comparison} compared the effectiveness of different CNN architectures in classifying benign and malignant breast masses from ultrasound images. The two CNN architectures used were VGG-16 \cite{simonyan2014very} and Inception-V3 \cite{szegedy2016rethinking}. Two training strategies were evaluated: using pre-trained models as feature extractors and fine-tuning pre-trained models. The dataset comprised 947 ultrasound images, including 587 images of benign masses and 360 images of malignant masses. Performance metrics used were accuracy and AUC (Area Under the ROC Curve). The results showed that fine-tuning VGG-16 achieved the best performance with an accuracy of 0.919 and AUC of 0.934. The comparison between the two training strategies indicated that fine-tuning the model generally outperformed using feature extraction.

    Several data augmentation methods have also been implemented by Walid Al-Dhabyani et al. \cite{al2019deep} to explore the use of deep learning in breast ultrasound image-based breast tumor classification. Specifically, the study focuses on two main aspects: identifying the significant problem of insufficient large datasets, which diminishes the performance of classification models. To address this challenge, the research proposes the use of data augmentation methods, including flipping, rotation, and adding noise, aiming to create a diverse and larger training dataset. In the realm of deep learning classification, the study concentrates on various architectures such as CNN, ResNet, and DenseNet to determine the benign or malignant nature of breast tumors. These architectures represent advancements in the field of deep learning and are effectively applied to medical image classification tasks. Additionally, the research investigates the performance of transfer learning models like VGG16 and MobileNet. Utilizing these models can leverage knowledge previously learned from large datasets, enhancing the classification ability of the model for breast ultrasound images. Among the studied architectures, DenseNet achieves the highest accuracy with an AUC of 0.976. In the study by Behnaz Gheflati et al. \cite{gheflati2022vision}, the authors employed the Vision Transformer (ViT) architecture for breast ultrasound image classification. The results indicate that ViT achieves a classification accuracy of 0.79, with an AUC of 0.84, outperforming ResNet. UGGNet is constructed upon the architecture of U-Net, where the encoder is substituted with convolutional blocks from VGGNet. The study draws inspiration from the Capsule Network (CapsNet) model, integrating it with the U-Net architecture, resulting in a hybrid model termed CapUnet \cite{guo2020deep}. This integration enables UGGNet to acquire intricate features from breast images while simultaneously retaining the classification capabilities inherited from VGGNet.

\section{Proposed method}

    \begin{figure*}[h]
        \begin{center}
            \includegraphics[scale=0.34]{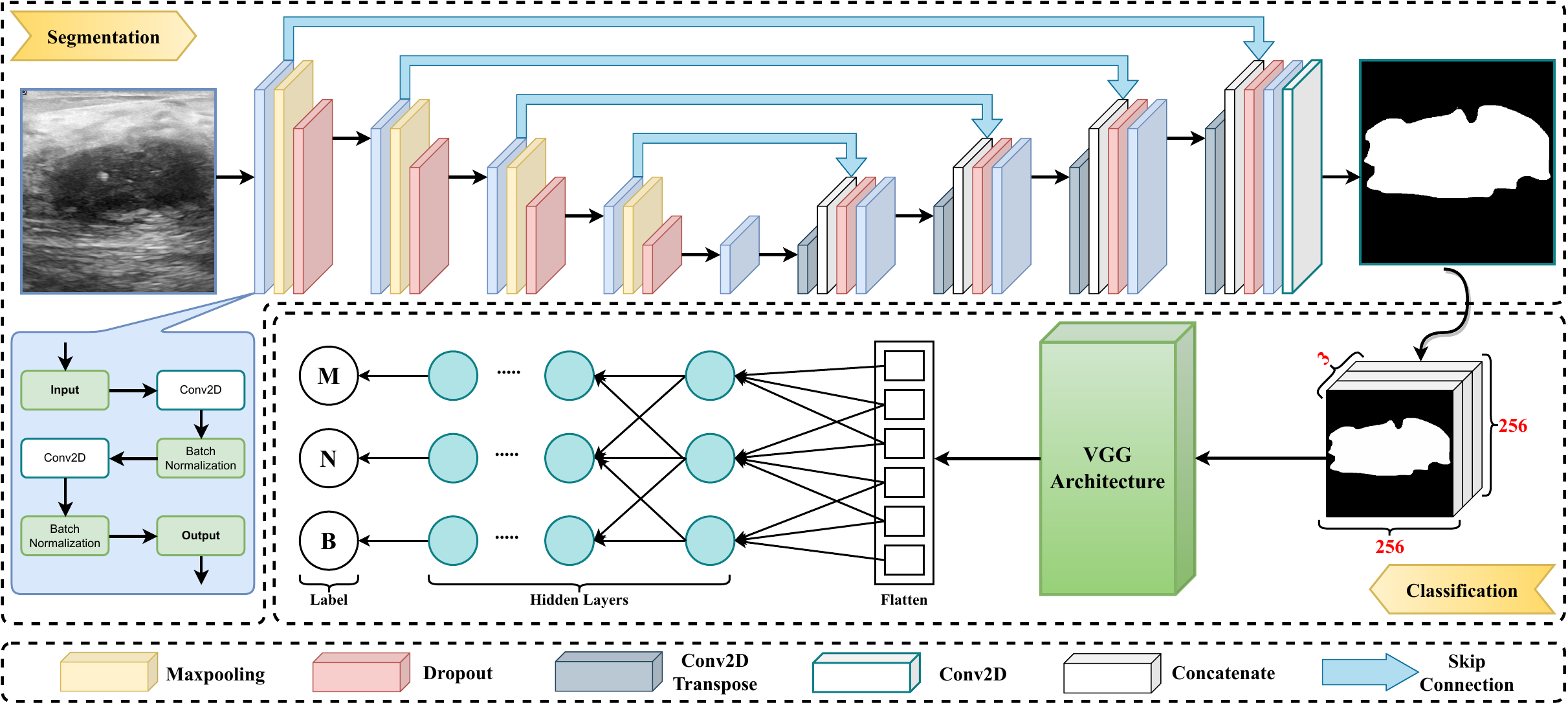}
        \end{center}
    \caption{Proposed UGG Architecture}
    \label{fg:Proposed UGG architecture}
    \end{figure*}
    
    The architecture UGGNet (Fig \ref{fg:Proposed UGG architecture})  consists of two main components: Segmentation with U-Net and Classification with VGGNet. The encoder process includes the following steps: the input image, with dimensions $256 \times 256 \times 3$, undergoes a custom-defined Conv2D-Block (shown in blue), which consists of two Conv2D layers and two BatchNormalization layers alternately. Subsequently, a Max\_Pooling layer (shown in yellow) with a kernel size of $2 \times 2$ follows the Conv2D-Block. At this point, the image is reduced in size by half compared to the image output from the Conv2D-Block, while the depth of the image is doubled. Assuming the previous image had dimensions $256 \times 256 \times 16$, the resulting image will have dimensions $128 \times 128 \times 32$. Next is a block (shown in red) with a 0.3 scaling factor, representing a Dropout layer. This process is repeated four times, starting from the input image $256 \times 256 \times 3$ and going through four layers, each defined as En\_L\_ (described in table \ref{tab:shape_Unet}). The layer specifications are as follows:

    \begin{table}[H]
        \centering
        \setlength\tabcolsep{15.1pt}
        \caption{Image shape during segmentation process}
        \label{tab:shape_Unet}
        \begin{tabular}{ccc}
            \hline
            \hline
            \textbf{No.} & Layer & Shape   \\
            \hline
            \hline
            1 & Input      & (256 x 256 x 3)  \\
            2 & En\_L\_1     & (128 x 128 x 16) \\
            3 & En\_L\_2     & (64 x 64 x 32)   \\
            4 & En\_L\_3     & (32 x 32 x 64)   \\
            5 & En\_L\_4     & (16 x 16 x 128)  \\
            6 & Last Image & (16 x 16 x 256)  \\
            7 & De\_L\_1     & (32 x 32 x 128)  \\
            8 & De\_L\_2     & (64 x 64 x 64)   \\
            9 & De\_L\_3     & (128 x 128 x 32) \\
            10 & De\_L\_4     & (256 x 256 x 16) \\
            11 & Out Image  & (256 x 256 x 1) \\
            \hline
            \hline
        \end{tabular}
    \end{table}

    The "Last Image" is the final image after the Max\_Pooling process in the 4th layer. At this point, the image contracts to dimensions $16 \times 16 \times 256$, significantly increasing the depth to 256 to capture as many features as possible. Following this, the decoder process begins: the "Last Image" is decoded and its dimensions increased by a factor of 2 using Conv2DTranspose. Similar to the encoder, the decoder consists of four layers defined as "De\_L\_". The output image "Out Image" of the 4th decoder layer is an image with dimensions $256 \times 256 \times 1$. This image is referred to as the mask image in the segmentation process, signifying the completion of the segmentation process.

    Following segmentation is the classification process. The image with dimensions $256 \times 256 \times 1$ cannot be directly input into the VGG architecture because the input requires dimensions of ($width,\ height,\ 3$). Therefore, the "Out Image" is passed through a Conv2D layer with a filter size of 3, which increases the depth of the image from $256 \times 256 \times 1$ to $256 \times 256 \times 3$. This image is then used as input for the VGG architecture. Subsequently, VGG produces output that stops at the layer \( \text{Conv5-4} \) for VGG19 and \( \text{Conv5-3} \) for VGG16. However, this output is not yet suitable for prediction.

    At this point, a Flatten layer is needed to "flatten" the tensor into a 1D vector with a shape of \( \left(\text{batch\_size},\ 512\right) \). This 1D vector is then fed into fully connected layer, and its output passes through a softmax activation function with three elements corresponding to the three labels (normal - N), (benign - B) and (malignant - M).
    
\section{Experiments}
    \subsection{Dataset}
    The dataset used in this study is named "Dataset of Breast Ultrasound Images" \cite{al2020dataset}, collected using the LOGIQ E9 ultrasound system \cite{deng2022engan}. It comprises 780 ultrasound images of female breasts, data was collected in 2018, focusing on the age group ranging from 25 to 75 years old. In total, this dataset comprises visual information of 600 female patients. The dataset is categorized into three labels: normal, benign, and malignant. Ultrasound images are associated with a corresponding mask image used for segmentation tasks. The average image size of the entire dataset is 500×500 pixels, stored in PNG format. 80\% will be used to train the UGGNet model, the remaining 20\% will be used for the final test set.
    
    \begin{figure}[h]
        \centering
        \includegraphics[width=\columnwidth]{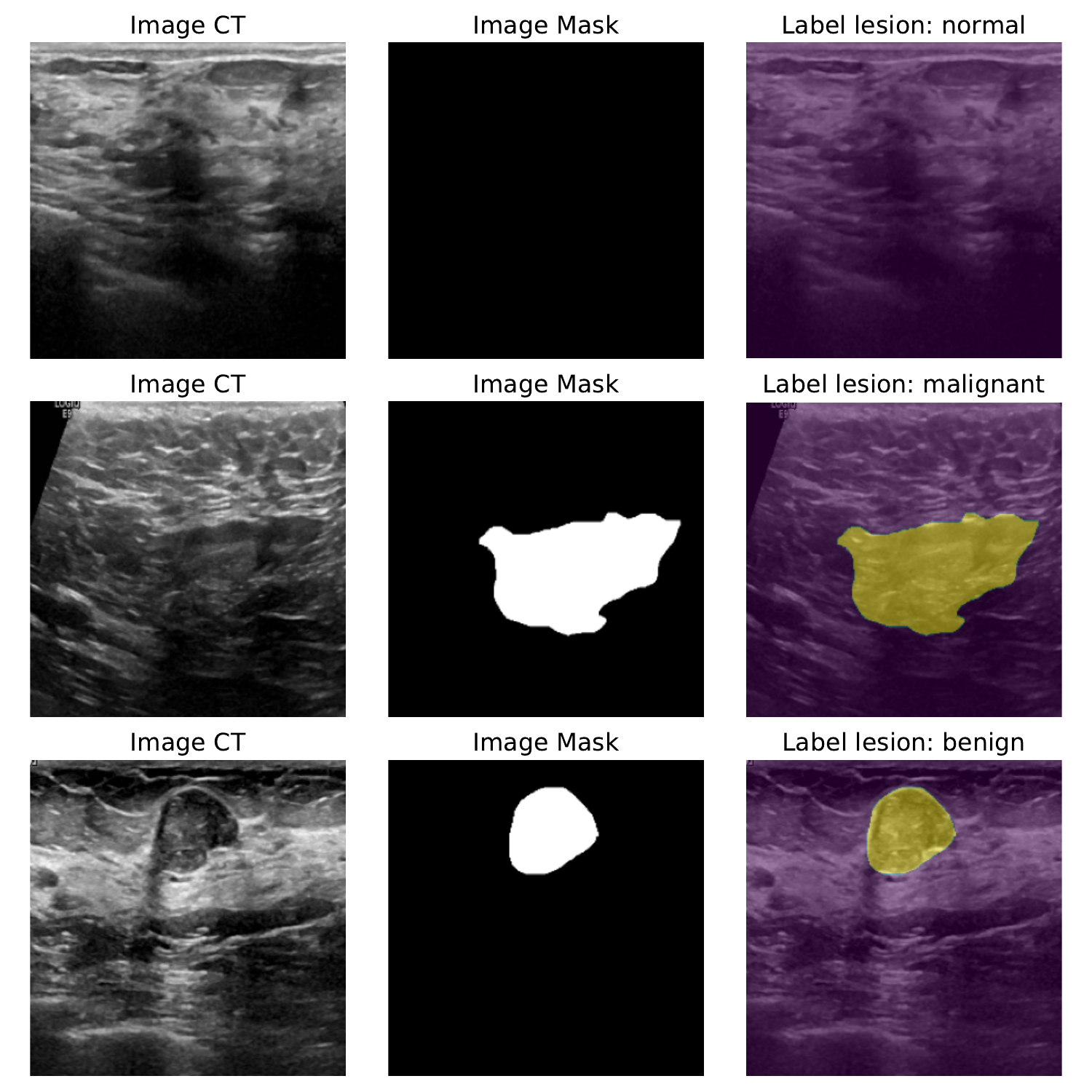}
        \caption{Ultrasound image with disease severity label}
        \label{fg:Ultrasound}
    \end{figure}

    \subsection{Metrics}
        \subsubsection{Loss function}
        The categorical cross-entropy loss function, denoted as \( H(y, \hat{y}) \), is a method for measuring the difference between the actual distribution \( y \) (true labels) and the predicted distribution \( \hat{y} \) (predicted probabilities) in a classification problem. The formula for the loss function is expressed as follows:
    
        \begin{equation}
        H(y, \hat{y}) = - \sum_{i}^{N} y_i \cdot \log(\hat{y}_i) 
        \end{equation}
    
        Here, \( y \) represents the actual distribution, where \( y_i \) is the probability of class \( i \) in the actual distribution. Conversely, \( \hat{y} \) is the predicted distribution, and \( \hat{y}_i \) is the predicted probability for class \( i \). Through the formula, each element \( y_i \) combined with the natural logarithm of \( \hat{y}_i \) is used to measure the discrepancy between the actual label and the corresponding prediction.
    
            

        \begin{algorithm}
            \caption{Categorical Cross-Entropy Loss Calculation}
            \label{alg:categorical_cross_entropy}
            \KwData{Actual distribution $y$, Predicted distribution $\hat{y}$}
            \KwResult{Categorical Cross-Entropy Loss $H(y, \hat{y})$}
            
            \BlankLine
            \textbf{Initialization:} Set $H(y, \hat{y})$ to 0\;
            
            \BlankLine
            \ForEach{class $i$}{
            $H(y, \hat{y}) \leftarrow H(y, \hat{y}) - y_i \cdot \log(\hat{y}_i)$\;
            }
            
            \BlankLine
            \textbf{Output:} Categorical Cross-Entropy Loss $H(y, \hat{y})$\;
        \end{algorithm}
         The detailed algorithm is presented in Algorithm \ref{alg:categorical_cross_entropy}. The goal is to minimize the value of the loss function, which is synonymous with making the predicted distribution close to the actual distribution. 
        This is an important tool in the process of training machine learning models, helping to shape and update the model's weights to optimize classification performance.

        \subsubsection{Evaluation Metrics}
        To assess the performance of a classification model, we use a Confusion Matrix. It operates on test data by categorizing predictions into four main types: True Positive (TP), True Negative (TN), False Positive (FP), and False Negative (FN).
        
            \begin{table}[h]
                \caption{Confusion Matrix}
                \begin{tabular}{|l|ll|}
                \hline
                \textbf{}                          & \multicolumn{2}{c|}{True Class}                                    \\ \hline
                \multirow{2}{*}{Predicted   Class} & \multicolumn{1}{l|}{True Positives}  & False Positives \\ \cline{2-3} 
                                                   & \multicolumn{1}{l|}{False Negatives} & True Negatives \\ \hline
                \end{tabular}
            \end{table}
        
            True Positive represents the number of cases where the model correctly predicts positive outcomes. True Negative is the number of cases where the model correctly predicts negative outcomes. False Positive is the number of cases where the model incorrectly predicts positive outcomes and False Negative is the number of cases where the model incorrectly predicts negative outcomes.
        
            Based on these four values, we can calculate various important metrics to evaluate the performance of the model, such as Recall, Accuracy, and F1-score.

            \begin{equation}
                \text{Accuracy} = \frac{TP + TN}{TP + TN + FP + FN}
            \end{equation}
            Accuracy is the percentage ratio of the number of correct predictions to the total number of data points. It provides an overall measure of the model's accuracy.

            \begin{equation}
                \text{Recall} = \frac{TP}{TP + FN}
            \end{equation}
            Recall measures the model's ability to correctly identify cases that truly belong to a specific class. The Recall formula is the number of correct predictions divided by the total number of cases that belong to that class.
            
            \begin{equation}
                \text{F1 Score} = 2 \times \frac{\text{Precision} \times \text{Recall}}{\text{Precision} + \text{Recall}}
            \end{equation}
            F1-score is a combination of Precision, Recall. Precision is the model's ability to make accurate predictions for cases predicted to belong to a specific class.

    \subsection{Implementations}
        There are two main versions of UGGNet, namely UGG\_19 and UGG\_16. UGG\_19 utilizes the U-Net architecture for feature extraction and employs VGG19 for classification, whereas UGG\_16 also utilizes U-Net but performs classification using VGG16.
        

        Training Parameters: During the training process of the machine learning model, the research utilized 80\% of the data for training and 20\% for model validation to ensure accuracy and performance. To initiate the training process, a learning rate (lr) of 0.0001 was employed, and a learning rate reduction was applied every 20 epochs if the validation accuracy (val\_accuracy) did not improve. The learning rate decay factor (Factor\_Decay\_Lr) was set to 0.8.
        
        For the training phase, a batch size (Batch\_Size) of 64 was selected to optimize model weight updates. Additionally, to prevent overfitting, dropout with a rate of 0.3 was applied to avoid excessive learning of specific features in the training data. The training process iterated over 500 epochs; however, to avoid resource consumption without significant improvement, Early Stopping was employed. The model would stop training after 100 epochs if the validation accuracy did not increase. To evaluate the model's performance, 20\% of the training data was used as a validation set (Validation\_Split). This approach helped control the training process and assess the model more generally.
        
        Finally, for model optimization, the research chose "adam" as the optimizer and "categorical\_crossentropy" as the loss function to ensure effective learning and the best performance measurement on the test set in the field of computer vision and deep learning.
        \begin{table*}[h]
            \centering
            \caption{UGGNet model size}
            \label{tab:UGGNet model size}
            \resizebox{\textwidth}{!}{%
                \begin{tabular}{cccccc}
                \hline
                \hline
                No. & \multicolumn{1}{c}{Model} & Num layer & Unit on layer & Total parameters & Training parameters \\
                \hline
                \hline
                1 & UGG\_16 & 3 & 1024/512/256              & 18.061.498 & 3.343.866 \\
                2 & UGG\_16 & 5 & 1024/512/256/128/64       & 18.102.074 & 3.384.442 \\
                3 & UGG\_16 & 7 & 1024/512/256/128/64/32/16 & 18.104.538 & 3.386.906 \\
                \hline
                \hline
                4 & UGG\_19 & 3 & 1024/512/256              & 23.371.194 & 3.343.866 \\
                5 & UGG\_19 & 5 & 1024/512/256/128/64       & 23.411.770 & 3.384.442 \\
                6 & UGG\_19 & 7 & 1024/512/256/128/64/32/16 & 23.414.234 & 3.386.906 \\
                \hline
                \hline
                \end{tabular}%
            }
        \end{table*}
        
        In the table \ref{tab:UGGNet model size}, UGG\_19 model, featuring a 7-layer architecture, appears significantly more complex compared to its predecessors. Designed with a total of 23,414,234 parameters, including 3,386,906 trainable parameters. 
        With units per layer sequentially set at 1024, 512, 256, 128, 64, 32, and 16, the augmentation of both the number of layers and units may imply enhanced learning capability and improved representation of complex data.
    \subsection{Experimental Results}
        The experimental setup will encompass six cases, incorporating UGG\_16 and UGG\_19 architectures, each characterized by the number of layers and units in the fully connected layers, specifically 3, 5, and 7 layers.
        \begin{table*}[h]
        	\centering
        	\caption{Experimental results of UGGNet model}
        	\label{tab:Experimental results of UGGNet model}
                \resizebox{\textwidth}{!}{
                \begin{tabular}{cccccccc}
                \hline
                \hline
                \textbf{No.} & Model   & Num layer & Epoch Stop & Training time & Accuracy & Recall & F1     \\
                    	\hline
                    	\hline
                1            & UGG\_16 & 3         & 292/500        & 01:24:14       & 0.7436   & 0.7436 & 0.739  \\
                2            & UGG\_16 & 5         & 239/500        & 01:10:03       & 0.7564   & 0.7564 & 0.7511 \\
                3            & UGG\_16 & 7         & 453/500        & 02:12:00      & 0.6923   & 0.6923 & 0.6991 \\
                \hline
                \hline
                4            & UGG\_19 & 3         & 229/500        & 01:11:11       & 0.7628   & 0.758  & 0.7628 \\
                5            & UGG\_19 & 5         & 178/500        & 00:54:41       & 0.7564   & 0.7564 & 0.7482 \\
                6            & UGG\_19 & 7         & 285/500        & 01:26:04       & 0.7821   & 0.7821 & 0.7754 \\
                \hline
                \hline
                \end{tabular}
                }
        \end{table*}
        In the table \ref{tab:Experimental results of UGGNet model} details of two models, UGG\_19 and UGG\_16, based on the number of layers. For UGG\_19, training with 3, 5, and 7 layers concluded at epochs 229, 178, and 285, respectively. The corresponding accuracies were 0.7628, 0.7564, and 0.7821, recall rates were 0.7580, 0.7564, and 0.7821, and F1 scores were 0.7628, 0.7482, and 0.7754. For UGG\_16, training with 3, 5, and 7 layers ended at epochs 292, 239, and 453. The accuracies were 0.7436, 0.7564, and 0.6923, recall rates were 0.7436, 0.7564, and 0.6923, and F1 scores were 0.7390, 0.7511, and 0.6991. Training durations varied, ranging from approximately 00:54:41 to 02:12:00.

        Compared to 2020 studies, Lazo et al. \cite{lazo2020comparison} optimized Inception V3, achieving accuracies of 0.713 and 0.756. In 2021, Irfan et al. \cite{Irfan2021dilated} proposed the di-Cnn Model, combining Densenet201 with a 24-Layer CNN, yielding an accuracy of 0.7961, detailed in the table \ref{tab:Compare the results of UGGNet with previous research}
    
        \begin{table*}[h]
			 \centering
			 \caption{Compare the results of UGGNet with previous research}
			 \label{tab:Compare the results of UGGNet with previous research}
                \resizebox{\textwidth}{!}{
			 \begin{tabularx}{\textwidth}{cc>{\centering\arraybackslash}Xcc}  
					\hline
					\hline
					\textbf{No.} & Model  & Method & Training time & Accuracy \\
					\hline
					\hline
					1 & Inception   V3 \cite{lazo2020comparison}  & Feature extraction with CNN    & -  & 0.7131   \\
					2 & Inception V3  - Fine-tuning \cite{lazo2020comparison} & Feature extraction with CNN combined with hyperparameter optimization   & -  & 0.7561   \\
					\hline
					\hline
					3 & Di-CNN \cite{Irfan2021dilated} & DenseNet201+ CNN 24 layer & 197:03:28     & 0.7961   \\
					\hline
					\hline
					4 & UGG\_19& Feature extraction with U-Net and classification with VGG19 + 7 Layer & 01:26:04      & 0.7821   \\
					5 & UGG\_16& Feature extraction with U-Net and classification with VGG16 + 5   Layer & 01:10:03      & 0.7564  \\
					\hline
					\hline
			 \end{tabularx}
                }
        \end{table*}

        The provided table presents a comparative analysis of various deep learning models, with a specific emphasis on Di-CNN and UGG\_19. Di-CNN, which employs DenseNet201 in conjunction with a 24-layer CNN, distinguishes itself with an impressive accuracy of 0.7961. However, a notable drawback lies in its extensive training duration, requiring 197 hours, 3 minutes, and 28 seconds. In contrast, UGG\_19 adopts a different strategy, leveraging U-Net for feature extraction and utilizing VGG19 with an additional 7 layers for classification. Despite a shorter training period of 1 hour, 26 minutes, and 4 seconds, UGG\_19 achieves a competitive accuracy of 0.7821. Remarkably, UGG\_19 attains commendable results through the synergistic combination of U-Net and VGG19 with additional layers, underscoring the importance of thoughtful architectural selection. This highlights the effectiveness of UGG\_19 as a model that strikes a balance between accuracy and training efficiency.
    
\section{Conclusions}
    The UGGNet model is proposed for identifying features in medical images, leveraging a unique combination of the U-Net and VGGNet architectures to achieve high performance. There are two main versions of UGGNet: UGG\_19 and UGG\_16. UGG\_19 employs the U-Net architecture for feature extraction and VGG19 for classification, whereas UGG\_16 also utilizes U-Net but employs VGG16 for classification. Experimental results have demonstrated that UGG\_19 achieves impressive performance, with the highest accuracy reaching 0.7821. Both Recall and F1 score are also impressive, at 0.7821 and 0.7754, respectively. The model was trained in an impressive time span of 1 hour, 26 minutes, and 4 seconds. The results of UGG\_19 showcase an effective synergy between the detailed feature extraction capability of U-Net and the classification prowess of VGG19. This outcome underscores the strength of employing hybrid architectural approaches to create a practical and efficient model for identifying features in medical images.


\bibliography{references}
\bibliographystyle{plain}

\end{document}